\documentclass[10pt]{article}

\usepackage{amssymb}

\usepackage{color}

\newcommand{\sm}[1]{{#1}}  

\let\oldfootnotetext\footnotetext
\renewcommand{\footnotetext}[1]{%
  \oldfootnotetext{\parbox{0.8\textwidth}{#1}}%
}

\usepackage[letterpaper]{geometry}
\usepackage{hicss}
\usepackage{times}
\usepackage[none]{hyphenat}
\usepackage{url}
\usepackage{graphicx}
\usepackage{multirow}
\usepackage{amsmath}

\usepackage[hang, flushmargin]{footmisc} 

\usepackage{stfloats}
\usepackage{float}

\usepackage{setspace}

\title{\sm{An Econometric Analysis} of Large Flexible \sm{Cryptocurrency-mining Consumers in Electricity Markets}}

\author{Subir~Majumder \\
   Department of Electrical and Computer Engineering,\\ Texas A\&M University,\\ College Station, TX 77840\\
  {\underline{subir-em@ieee.org}} \\ \\
   \\ \And
   ~~\\ ~~\\ ~~\\ \\ \\
  Le~Xie \\
   School of Engineering and Applied Sciences,\\ Harvard University,\\ Allston, MA 02134 \\ 
  {\underline{xie@seas.harvard.edu} } \\ ~~\\ \And
  Ignacio~Aravena \\
  Lawrence Livermore National Laboratory,\\ Livermore, CA 94550 \\
  {\underline{aravenasolis1@llnl.gov} } \\ \\
   \\ } 

\date{}

\begin{document}
\maketitle
\begin{abstract}
In recent years, power grids have seen a surge in large cryptocurrency mining firms, with \sm{individual} consumption levels reaching 700MW. This study examines the behavior of these firms in Texas, focusing on how their consumption is influenced by cryptocurrency conversion rates, electricity prices, local weather, and \sm{other} factors. We transform the skewed electricity consumption data of these firms, perform correlation analysis, and apply a seasonal autoregressive moving average model for analysis. Our findings reveal that, surprisingly, short-term mining electricity consumption is not directly correlated with cryptocurrency conversion rates. Instead, the primary influencers are the temperature and electricity prices. These firms also respond to avoid transmission and distribution network (T\&D) charges --- commonly referred to as four Coincident peak (4CP) charges --- during the summer months. As the scale of these firms is likely to surge in future years, the developed electricity consumption model \sm{can be used to} generate public, synthetic datasets to understand the overall impact on the \sm{power grid}. The developed model could also lead to better \sm{pricing} mechanisms to effectively use \sm{the flexibility of these} resources \sm{towards improving} power grid reliability.
\end{abstract}

\subsubsection*{Keywords:}

Demand Response, Econometric Model, Large Flexible Cryptomining Loads, Electricity Markets.

\section{Introduction}

\begin{figure}[!]
    \centering
	\includegraphics[width=0.89\linewidth]{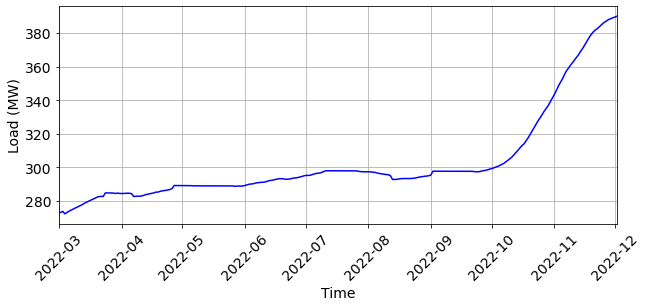}
	\caption{Trend of large cryptocurrency mining loads in a typical ERCOT load zone.}
	\label{fig:reg_trnd}       
\end{figure}

\sm{As shown in Fig. \ref{fig:reg_trnd} (which is corroborated by slide 3 of \cite{ercot2024_LLI}), the Texas electric grid is facing a rapid cryptocurrency mining data-center-driven load growth.} 
The Electricity Reliability Commission of Texas (ERCOT)\sm{---}the market operator \sm{in charge of the} largest part of the Texas electricity grid\sm{---}allows both generators and loads to truthfully disclose their price sensitivity and \sm{be} dispatched through their economic dispatch process. As evidenced through the U.S. Securities and Exchange Commission (SEC) annual report (available in \cite{riotblockchain2022}), some of these cryptocurrency mining firms participate in the ERCOT electricity market. However, \cite{ercot2022lfltf} observes \sm{that the cryptocurrency mining firms, }
with an individual capacity greater than or equal to 75.0 MW, 
show 
price inflexibility in their electricity consumption offer curve within a given settlement interval. These firms also exhibit price flexibility across multiple settlement intervals; for example, as shown in Table \ref{table:correlation1}, these firms significantly reduce their demand during summer months when high system-wide load stresses the power grid. \cite{ercot2024_LLI} also reports that the responsiveness of these firms is not uniform across multiple facilities when exposed to the same circumstances. On one hand, the challenges in ensuring grid reliability under these emerging circumstances can be concerning for any power grid operator facing a similar stream of cryptocurrency mining data-center interconnections. On the other hand, it is possible for the power grid operators to model these firms appropriately for harvesting flexibility from these resources for the overall benefit of the power grid.

\begin{table}[h!]
\caption{Day-time Correlation between Loads and Average Temperatures across Texas\\}\label{table:correlation1}
\resizebox{\columnwidth}{!}{%
\begin{tabular}{|c|c|c|}
\hline
          & \begin{tabular}[c]{@{}c@{}}Cryptocurrency-mining\\  Firms' Response\end{tabular} & \begin{tabular}[c]{@{}c@{}}ERCOT-wide Load \\ Response\end{tabular} \\ \hline
Non-summer   & -0.17                                                                  & 0.78                                                                \\ \hline
\sm{S}ummer       & -0.40                                                                  & 0.89                                                                 \\ \hline
\end{tabular}
}
\end{table}

In this article, our focus is on \textit{large cryptocurrency mining firms} with an individual capacity greater than or equal to 75.0 MW. These mining firms typically operate as part of a mining pool\footnote{Refer to: https://www.investopedia.com/terms/m/mining-pool.asp}, where the \sm{Bitcoin} reward is a function of the hashing power contributed. Hashing power is directly related to the energy consumption of mining loads. Therefore, The operating costs of these firms include electricity procurement through various \sm{mechanisms}, including long-term power purchase agreements, and \sm{transactions at} ERCOT's electricity market\sm{s}. These firms are also expected to share the cost burden of Texas's power grid infrastructure. In this regard, ERCOT employs a fixed-cost recovery mechanism, where it identifies the four highest 15-minute electricity usage intervals each month from June to September\sm{---}during peak demand times\sm{---}and proportionately allocates the \sm{fixed} transmission and distribution (T\&D) network costs among all load \sm{participants based on their average consumptions in these 4 intervals}. As discussed in \cite{du2019demand}, these usage intervals are calculated on an \textit{ex-post} basis, and T\&D costs divided by average peak-demand across ERCOT is defined as (4 coincident peaks) 4CP prices. As an example, for a 500 MW cryptocurrency-mining firm\sm{, operating at full-load during these peak intervals}, the annual fixed cost will be $500 \text{MW} \times \$4.96/\text{4CP kW} \times 1000 \times 12 = $ \$29.76M (the rates for 4CP charges are taken from \cite{oncor_tariff}), which will be a significant portion of the \sm{facility's operation} cost. From the SEC annual reports (see \cite{riotblockchain2022}), we note that these cryptocurrency mining firms earn power curtailment credits by participating in ERCOT's ancillary services markets and strategically reducing their energy consumption. 

Thus, the profit of a mining facility ($P^M$) can be defined as:
\vspace{-4pt}
\begin{equation}
P^M = \sum_{\forall t} \left( \pi^{B}_{t} k^B E^H_{t} - \pi^{R}_{t} E^R_{t} - \pi^{D}_{t} E^D_{t}  + \gamma(E^M_{t}) \right) \label{miner:profit}
\end{equation}
\vspace{-6pt}
\sm{where,}
\begin{equation}
E^M_{t} = E^P_{t} + E^D_{t} + E^R_{t} = E^H_{t} + \psi(E^H_{t}, T_{t})  \label{energy}
\end{equation}
In \eqref{energy}, we observe that the total electricity procured by a mining facility ($E^{M}_{t}$) is equal to the sum of electricity purchased through long-term power purchase agreements ($E^{P}_{t}$), and \sm{day-ahead ($E^{D}_{t}$), and real-time ($E^{R}_{t}$)} electricity markets. Of the total energy procured, the miners use a portion of their procured energy for hashing ($E^{H}_{t}$) and another part for cooling, which is a function $\psi(\cdot)$ of hashing power and ambient temperature ($T_{t}$). In \eqref{miner:profit}, $\pi^{B}_{t}$ represents the \$ exchange rate of the cryptocurrency, while $\pi^{D}_{t}$ and $\pi^{R}_{t}$ are the day-ahead and real-time electricity market prices, respectively\sm{, for interval $t$}. Parameter $k^B$ is the efficiency of cryptocurrency miners' power supply. The function $\gamma(\cdot)$ represents the opportunity cost by avoiding 4CP \sm{charges, which is a function of the miner's electricity consumption $E^M_t$}. Therefore, the miners' short-term net profit is the sum of their revenue from selling cryptocurrency, the cost of power procured from the electricity markets, and the avoided cost of not hashing. However, it is extremely difficult to solve this problem because of three challenges. First, aside from power purchase agreements, all the prices are only known on an \textit{ex-post} basis. Second, as we noted from the SEC annual report of one crypto mining firm (available in \cite{riotblockchain2022}), miners do not sell their entire cryptocurrency inventory, implying that the \sm{expected} value of holding cryptocurrency can be higher than the current exchange rate. Third, the 4CP avoidance cost for cryptocurrency miners can be extremely complex to compute. 

\begin{figure*}[!h]
    \centering
	\includegraphics[width=\linewidth]{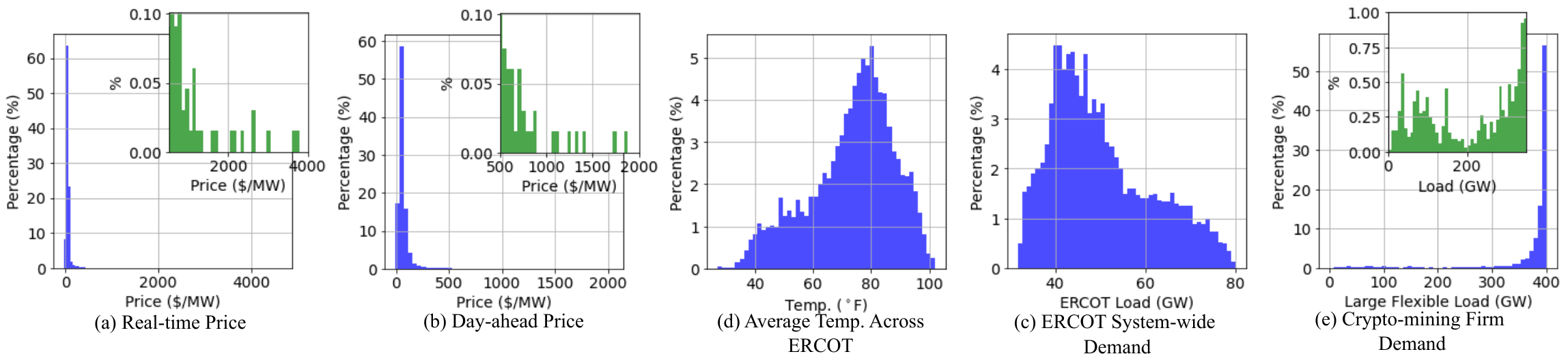}
	\caption{Histogram of the various hourly datasets for Apr.-Oct. 2022.}
	\label{fig:hist}       
\end{figure*}

The response of other industrial facilities to electricity prices has already been thoroughly discussed in the literature (see \cite{golmohamadi2022demand} for a recent review article). For example, in the aluminium smelting industry, \cite{depree2022contribution} have discussed `arbitrage price,' which identifies a correlation between electricity prices and aluminium prices. However, contrary to other industries, cryptocurrency mining firms are different in two ways. First, the exchange rate of cryptocurrencies is highly volatile. Second, cryptocurrencies can be stored in infinite quantities and \sm{for indefinite} periods. Therefore, cryptocurrency mining firms may not be subjected to the same market forces as in other industries. 
In regards to cryptocurrency mining firms, \cite{rhodes2021impacts} have already provided high-level behavioral analyses of cryptocurrency mining firms. \cite{menati2023modeling} have studied the impacts of various demand response programs for cryptocurrency mining loads in Texas. \cite{menati2024optimization} have also designed algorithms for miners to participate in the ERCOT market for profit maximization.

However, there is a lack of large-scale, data-driven analyses that provide predictive insights into why and to what extent cryptocurrency mining firms respond to various exogenous factors. To address these challenges, as highlighted in \eqref{energy}, we have regressed cryptocurrency mining firms' electricity consumptions against the ambient temperature, cryptocurrency prices, and day-ahead and real-time electricity prices. Since 4CP prices are based on ERCOT system-level electricity consumption during summer months, mining facilities may use this additional predictor to hedge against consumption during 4CP hours, which will also impact energy consumption. Additionally, there could be other endogenous factors based on historical operating experience that may not be explained by the exogenous factors discussed before.

Based on these insights, through a thorough data analysis, this paper proposes an autoregressive model with exogenous variables (AR-X) to identify cryptocurrency mining firm's flexibility in electricity consumption. \sm{We develop two} AR-X models, one describing the demand during summer and \sm{the other during} non-summer months. These developed models would not only be able to predict cryptocurrency mining firms' behavior but can be used to generate synthetic data for large-scale power system simulations under various environmental and market scenarios, helping electric energy system planners, market operators, and policymakers in the decision-making.

\begin{table*}[]
\caption{Summary Statistics for All Variables Contributing to Miners' Energy Consumptions}\label{table:correlation}
\centering
\resizebox{1.6\columnwidth}{!}{%
\begin{tabular}{|c|c|c|c|c|c|c|}
\hline \hline
\multirow{2}{*}{Metric} & Real Time & Day Ahead & ERCOT System & Texas Average & {Crypto-mining Firm} \\
& Price (\$/MWh) & Price (\$/MWh) & Demand (MW) & Temperature ($^{\circ}$F) & Demand (MW)\\ \hline \hline
Mean & 65.31 & 68.48 & 50218.1 & 73.4 & 370.59 \\ \hline
Standard Deviation & 136.56 & 79.5 & 11168.27 & 14.58 & 70.67 \\ \hline
Skewness & \textbf{18.58 } & \textbf{10.52 } & 0.67 & -0.6 & \textbf{-3.57 } \\ \hline \hline
J-B Test p-Values & \textbf{0 } & {\textbf{0} } & \textbf{0} & \textbf{0} & \textbf{0} \\ \hline
ADF Statistic p-Values & 0 & 0 & 0.03 & 0.04 & 0 \\ \hline
BP Test p-Values & 0.47 & 0.44 & \textbf{0.01 } & 0.79 & \textbf{0 } \\ \hline
Durbin Watson Test & \textbf{0.26 } & \textbf{0.12 } & \textbf{0 } & \textbf{0 } & \textbf{0 } \\ \hline \hline
\end{tabular}
}
\end{table*}

\section{Exploratory Data Analysis}


Fig. \ref{fig:hist} depicts the histogram of the hourly time-series panel data for these cryptocurrency mining firms' (electricity) demand and related explanatory parameters from March to October 2022. The electricity price data includes average real-time and day-ahead prices across all ERCOT load zones. ERCOT system demand represents the aggregated electricity demand across all ERCOT-managed regions in texas\sm{\footnote{All these datasets are sourced from \url{www.ercot.com}.}}. For average temperature, we collected weather data from several weather stations across ERCOT-managed regions in Texas\sm{\footnote{Available from \url{www.wunderground.com}.}}, and calculated the average temperature across these stations. The crypto-mining firms' electricity consumption dataset represents hourly load data aggregated across an ERCOT load zone, is not publicly available, and can be obtained upon request. This electricity consumption is mixed with other firm load data, which is unknown to us but is relatively small compared to crypto-firm consumptions.

As shown in Fig. \ref{fig:hist} with summary statistics detailed in Table \ref{table:correlation}, crypto-mining firm electricity demand and prices exhibit significant skewness. From the p-values of the Breusch-Pagan (BP) test, the dataset, particularly cryptocurrency miners' electricity consumption and electricity prices, displays non-constant variance (heteroscedasticity). If not addressed, this skewness and heteroscedasticity can cause inaccuracies in regressive models. As discussed in \cite{IntroductoryEconometrics7th}, according to the Gauss-Markov assumption, the error terms must have zero conditional means and be homoskedastic for linear regression estimators to remain unbiased. Additionally, ensuring normality in the error distribution is essential for applying the Central Limit Theorem, which aids in inferential statistics, including hypothesis testing and constructing confidence intervals.


\begin{figure*}[!h]
    \centering
	\includegraphics[width=0.90\linewidth]{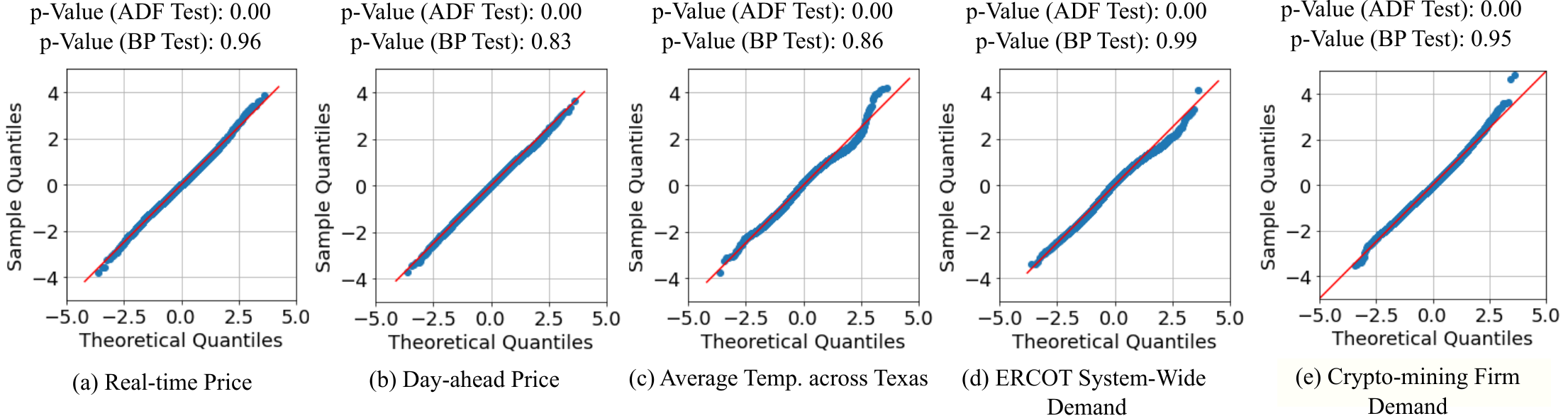}
	\caption{Q-Q plots for transformed datasets.}
	\label{fig:qq-transformd}       
\end{figure*}

\subsection{Data Transformation}

While performing the time-series analysis, it is essential to remove all diurnal and seasonal patterns in the datasets. As discussed by \cite{KRZYSZTOFOWICZ1997286}, transformations are applied to achieve an approximately Gaussian distribution, especially in those cases where the panel data are heavily skewed. While this transformation is not mandatory, it helps in ensuring that the residuals satisfy the Central Limit Theorem \sm{(i.e., namely that a model constructed from sequential addition of random variables will, under mild assumptions, inevitably exhibit Gaussian characteristics)}. There is no specific sequence for applying these steps. Given the exponential growth in crypto-miners penetration in the ERCOT grid, as highlighted through Fig. \ref{fig:reg_trnd}, we first extract responsive components from the general trend. Here, we assumed that the daily peak mining load demand remains constant within a rolling window, which also provides us with the trend component. The hourly time-series miners' consumption is obtained by dividing the trend component from the actual time-series data. The transformation and standardization steps are given below:



\begin{itemize}
    
\item[i.] We apply a non-parametric transformation to make the dataset, $y_{s,d}$, approximately follow a Gaussian distribution. The inverse quantile transform, a non-parametric technique, sorts the dataset in monotonic order, estimates cumulative probabilities, and identifies discrete quantiles for transformation. The transformation process is as follows:
\vspace{-2pt}
\begin{equation}
y^{G}_{s,d} = Q^{-1}\left( y_{s,d} \right)
\end{equation}
\vspace{-6pt}
\item[ii.] We remove seasonality and diurnal effects by normalizing the dataset using the sample mean and standard deviation:
\vspace{-2pt}
\begin{equation}
\Tilde{y}_{s,d} = \frac{y^{G}_{s,d} - \hat{\mu}_{s,d} }{\hat{\sigma}_{s,d}}
\end{equation}
\vspace{-6pt}
\end{itemize}
The quantile plots, Augmented Dickey-Fuller (ADF) statistic p-values, and BP test p-values in Fig. \ref{fig:qq-transformd} show that all transformed datasets are normally distributed, stationary and homoskedastic.

\subsection{Correlation Analysis}
\vspace{-20pt}
\subsubsection{Value of cryptocurrencies}

\begin{figure}[b!]
    \centering
	\includegraphics[width=\linewidth]{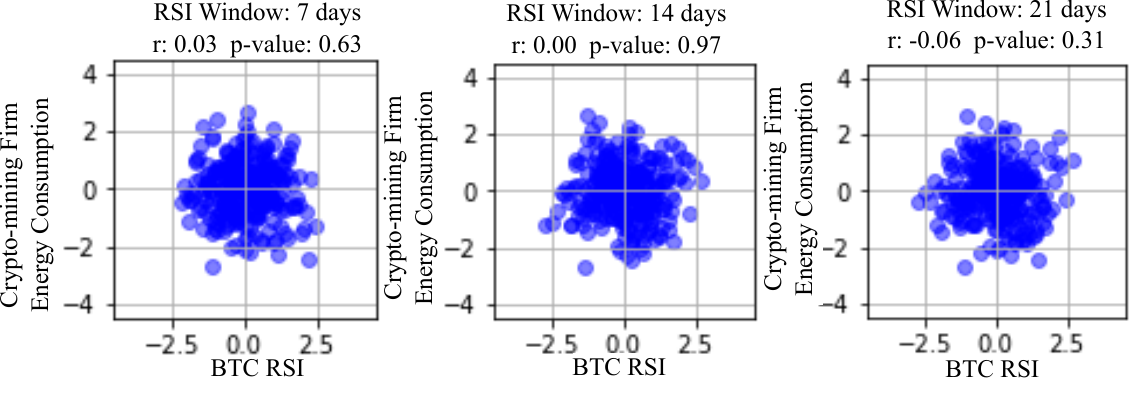}
	\caption{Comparing RSI of \sm{Bitcoin} and daily energy consumption of crypto-mining firms.}
	\label{fig:FLEvsBTC}       
\end{figure}

\begin{figure}[b!]
    \centering
	\includegraphics[width=0.91\linewidth]{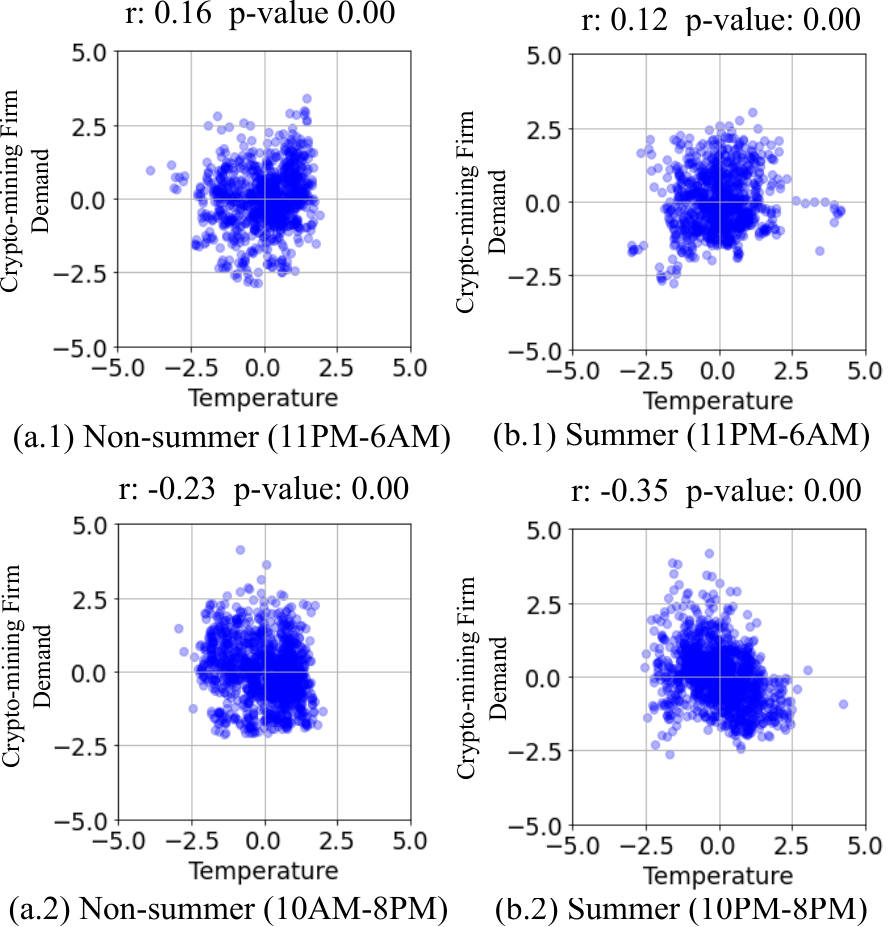}
	\caption{Correlation identifying how much of crypto-miners electricity consumption is responsible for cooling.}
	\label{fig:temp}       
\end{figure}

\begin{figure*}[b!]
    \centering
	\includegraphics[width=0.85\linewidth]{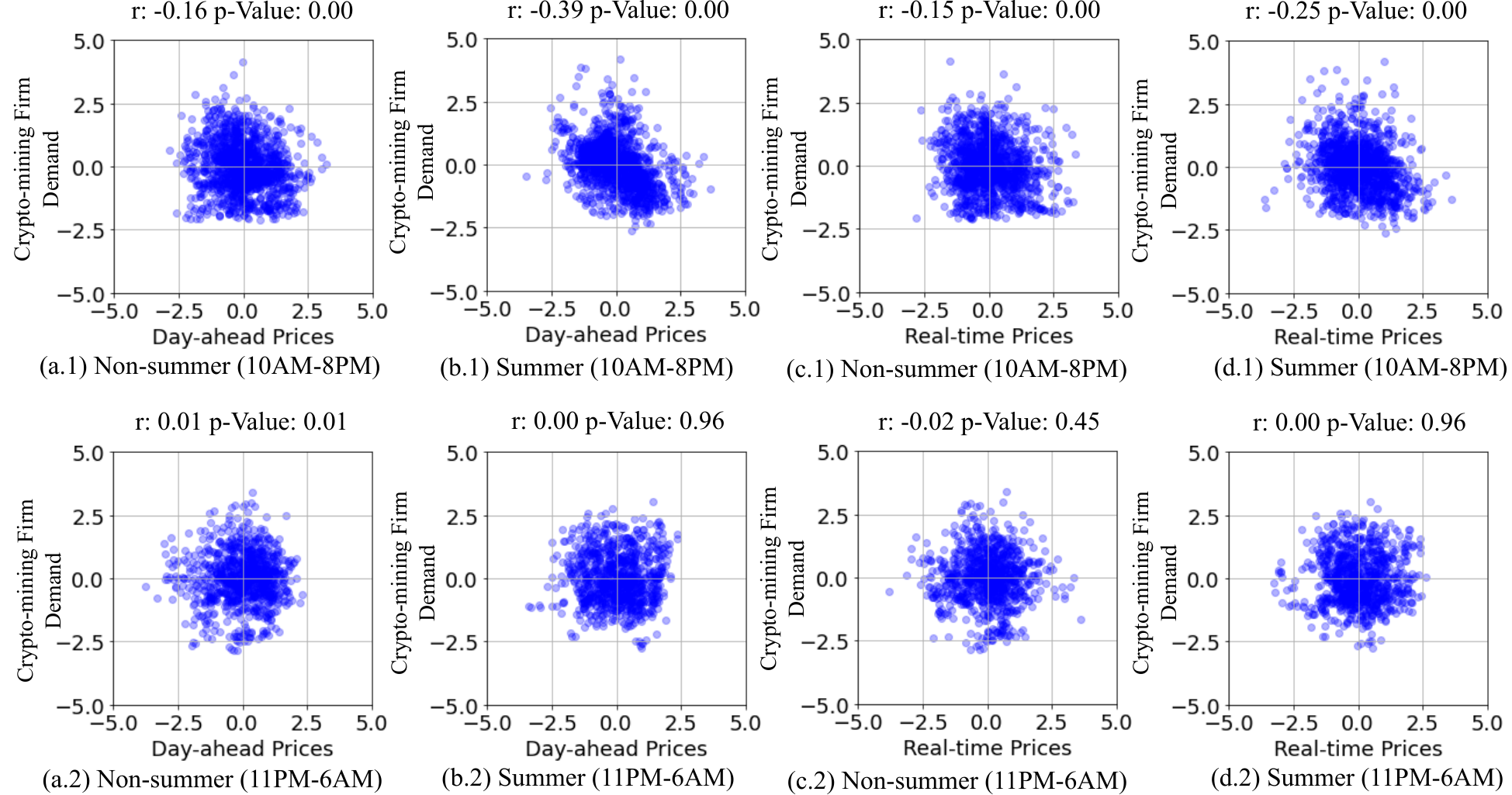}
	\caption{Cryptocurrency mining firms responses to prices.}
	\label{fig:prices}       
\end{figure*}

The dependence of the bitcoin exchange rate on energy consumption has been discussed in \cite{menati2023modeling}, which motivated us to perform this analysis. First, we only have access to historical daily Bitcoin exchange rate data, making it difficult to compare it against hourly cryptocurrency miners' electricity consumption. Secondly, the panel data is for the year 2022, when Bitcoin prices generally exhibited a downward trend, while, as shown in Fig. \ref{fig:reg_trnd}, there is an overall upward trend in cryptocurrency miners' daily peak electricity consumption. This could lead to incorrect conclusions about the relationship between cryptocurrency miners' energy consumption and Bitcoin exchange rate. We focus on bitcoin mining firms here because of their high energy intensity. To address these limitations, firstly, we calculated the daily net energy consumption for miners using detrended electricity consumption data. Secondly, instead of using actual Bitcoin prices, as identified by \cite{levy1967relative}, we employed the Relative Strength Index (RSI) --- a momentum measure describing the speed and magnitude of a security's price changes. The RSI is a short-term measure of overvalued or undervalued security conditions. We wanted to investigate if the crypto miners are using indices similar to RSI to control their daily electricity consumption.

The scatter plots of the RSI of the Bitcoin exchange rate considering the 7, 14, and 21-day window and the daily energy consumption of crypto-mining firms, depicted in Fig. \ref{fig:FLEvsBTC}, show p-values of correlation coefficient of $\geq$ 0.05 in all three cases. This suggests that, given the panel data concerned, cryptocurrency miners are agnostic to Bitcoin prices in the short term. However, mining facilities could utilize factors that are functions of Bitcoin exchange rates, which our simple correlation analysis could not capture. For example, as shown in \cite{ercot2024_LLI}, ERCOT utilizes strike prices to determine the responsiveness of cryptocurrency facilities, but it is not quite clear if those prices were generated based on current Bitcoin exchange rates. 




\subsubsection{The cooling energy requirements}

A significant portion of the energy consumed by cryptocurrency mining firms is dedicated to cooling (see eq. \sm{\eqref{energy}}). The cooling requirements are influenced by factors such as ambient temperature, the efficiency of the cryptocurrency miners, and hashing energy consumption. In our case, we are considering aggregated electricity consumption data across multiple crypto-ming firms, and we are not aware of the locations of mining firms, which is why, in this article, we consider average Texas temperature as a predictor. We observe, during the daytime, the strong correlation between temperature and system-wide electricity prices can obscure the cooling energy consumption. Even during non-summer months, temperatures \sm{can} remain high into the late evening. As illustrated in Fig. \ref{fig:temp}, from 10 PM to 6 AM, both in non-summer and summer periods, we observe weak positive correlations between electricity consumption and temperature with p-values close to 0. This confirms the physical principle that higher ambient temperatures necessitate more electricity for cooling.


\vspace{-5pt}
\subsubsection{Price responses}

If we ignore a few price peaks, historically in the ERCOT market\sm{---}as shown in Fig. \ref{fig:hist}(a,b)\sm{---}day-ahead prices are statistically higher than real-time prices and have a comparatively narrower standard deviation. This implies that day-ahead prices remain elevated for longer periods. Therefore, cryptocurrency miners' response to day-ahead prices will be stronger than their response to real-time prices, especially during the summer. Prices tend to be statistically lower at night, suggesting that cryptocurrency miners may not be incentivized to respond to either day-ahead or real-time prices during both summer and non-summer months during late-night hours. During the summer, prices remain higher than during non-summer months, as shown in Fig. \ref{fig:monthly}. We observe that cryptocurrency miners respond more vigorously to both day-ahead and real-time prices during the summer months. These price-responsive behaviors are depicted in Fig. \ref{fig:prices}. While not shown for brevity, cryptocurrency miners respond further vigorously during peak demand hours (3 PM-7 PM). The correlation coefficient for day-ahead prices during non-summer times increases to -0.29 (p-value 0.00), and during summer times to -0.42 (p-value 0.00). However, selecting a narrower window for real-time prices did not significantly increase the correlation coefficients.


\subsubsection{The predictors contributing to 4CP responses}

\begin{figure}[!]
    \centering
	\includegraphics[width=0.92\linewidth]{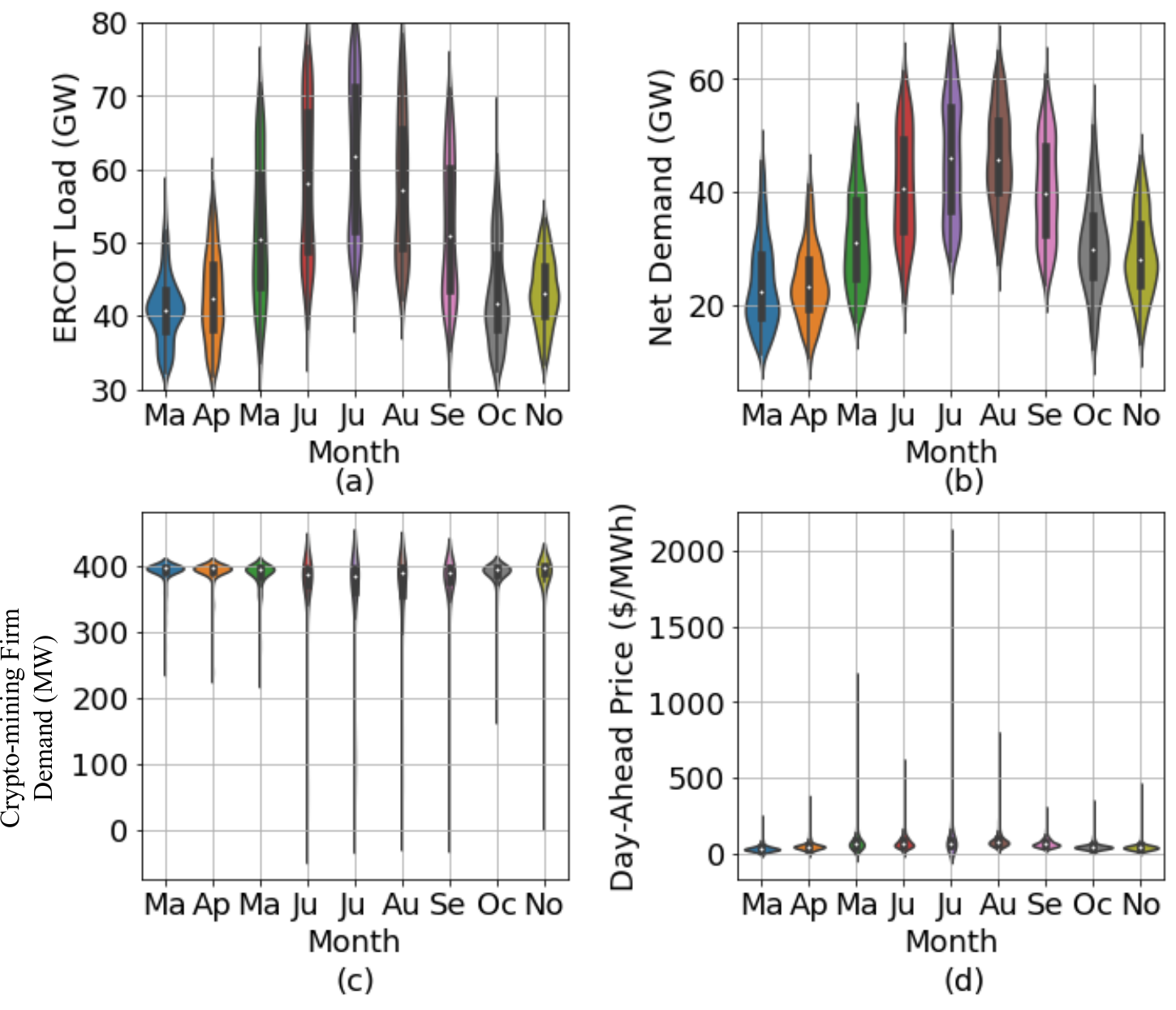}
	\caption{Comparing Demand across ERCOT, net-load, mining firm consumption, and day-ahead price.}
	\label{fig:monthly}       
\end{figure}

There are three main issues with using simple price-correlation to understand the direct impact of \sm{electricity} prices on crypto-mining firms' electricity consumption. Firstly, prices are not known \textit{a priori}. Consequently, cryptocurrency miners must decide whether and how much to shut down their facilities' latest \sm{with} the real-time market at least one hour in advance because of market rules. Secondly, as observed in Fig. \ref{fig:monthly}(c-d), the day-ahead prices in June, August, and September were not significantly higher, yet cryptocurrency miners responded as vigorously as they did in July. Thirdly, as shown in Fig. \ref{fig:prices}, when electricity prices are low, it is trivial for miners to operate at full capacity. This implies that cryptocurrency miners are likely using factors other than electricity prices to control their energy consumption during summer months, which must be to avoid 4CP charges.

\begin{figure}[!]
    \centering
      \includegraphics[width=0.94\linewidth]{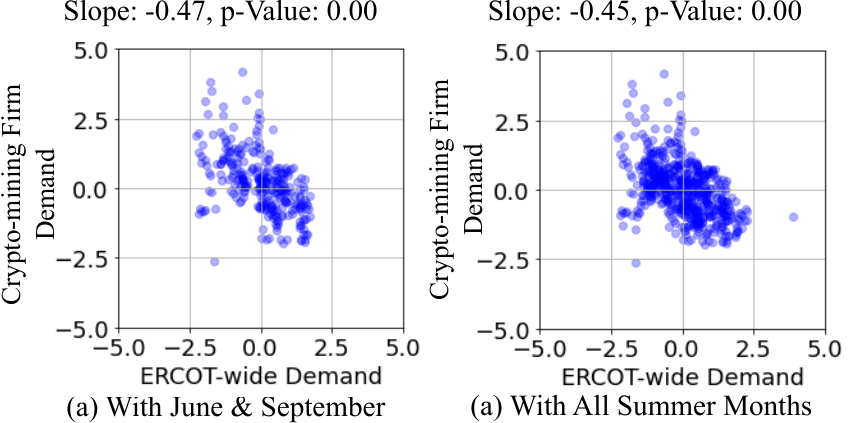}
	\caption{Factors responsible for 4CP demand response in addition to prices.}
	\label{fig:4CP}       
\end{figure}

4CP peaks are calculated based on ERCOT-wide demand and are price-agnostic. For example, in August 2023, the peak demand occurred on the 10th, while the price peaked at approximately \$4000/MWh on the 11th. Except for a few instances, 4CP peaks in ERCOT generally arise between 4 PM and 6 PM. There exists a challenge when regressing crypto-mining firm's electricity consumption against ercot system-wide loads in general because a higher load leads to higher electricity prices. To capture how the crypto-mining firms are hedging against 4CP prices, we need to focus on months when electricity prices were low, such as June and September (see Fig. \ref{fig:monthly}(d)), within hours 4 PM-6 PM. As depicted in Fig. \ref{fig:4CP}, the correlation between electricity consumption of crypto-mining firms and ERCOT system-wide electricity demand appears strongest when considering months with lower electricity prices alone. It is also notable that ERCOT records historical forecasts, which could be included as a predictor in future work.

\subsubsection{Auto-regressive Model}



The Durbin-Watson tests in Table \ref{fig:hist} indicate a significant presence of autocorrelation within the cryptocurrency miners' electricity consumption dataset. Autocorrelation occurs when variables are correlated with their own past values, suggesting that the electricity consumption of cryptocurrency mining facilities is influenced by their historical operational patterns. 

\begin{figure}[b!]
    \centering
	\includegraphics[width=0.95\linewidth]{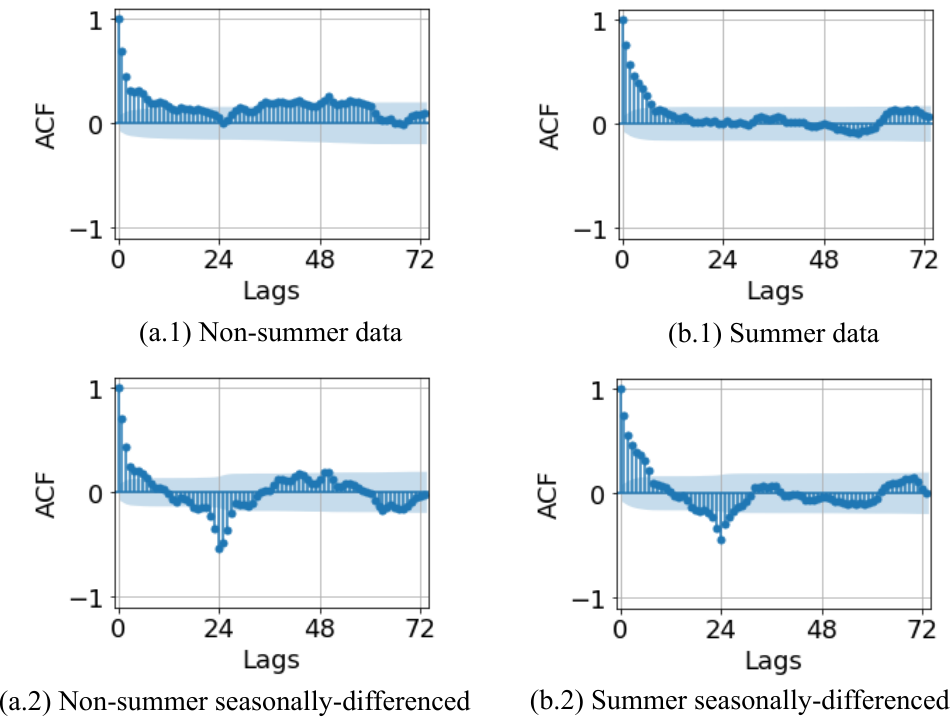}
	\caption{ACF plots without and with seasonal differencing considering both non-summer and summer months.}
	\label{fig:ACF-PACF}       
\end{figure}

Auto-Regressive Integrated Moving Average (ARIMA) processes are a class of stochastic processes used to analyze time series data. The ARIMA process, attributed to \cite{box2015time}, hypothesizes that the residual term is randomly drawn from a normal distribution with zero mean and constant variance, known as a white noise process. However, ARIMA models can be robust to the non-normality of residuals. As with other time-series analyses, the residuals need to be homoskedastic, and the time series itself must be stationary. A general ARIMA model is formally defined as follows:
\vspace{-3pt}
\begin{equation}
   \Phi(B^S) \phi(B) \nabla^d \nabla_s^D y_t = \Theta(B^S)\theta(B) \epsilon_t
\end{equation}
\vspace{-5pt}

\noindent where $y_t$ is the modeled cryptocurrency miners electricity consumption data. Here, $B$ is the backshift operator, where $B^l r_t: = r_{t-l}$, and $S$ is the seasonality of the time series. The functions representing auto-regressive, moving average, and differences and their seasonal forms are defined as $\phi(B) = 1 - \sum_{i=1}^{p} \phi_i B$, $\theta(B) = 1 + \sum_{i=1}^{q} \theta_i B$, $\nabla^d y_t = (1 - B)^d y_t$, $\Phi(B^S) = 1 - \sum_{i=1}^{P} \Phi_i B^S$, $\Theta(B^S) = 1 + \sum_{i=1}^{Q} \Theta_i B^S$, and $\nabla_S^D y_t = (1 - B^S)^D y_t$. The parameters $p,d,q,P,D,Q$ and $S$ identify the specific ARIMA process.

Autocorrelation factors (ACF) for both non-summer and summer months are plotted in Fig. \ref{fig:ACF-PACF}. Spikes around a lag of 24 in the ACF plots, which become prominent with seasonal differencing of 24 periods, suggest that the data exhibits seasonality, which is expected since the time-series dataset is hourly. The diminishing ACF plots indicate the presence of moving average components. 



\vspace{-7pt}
\setlength{\jot}{-3pt} 
\section{Empirical Observation of Cryptocurrency Mining Firms' Behavior}
\vspace{-7pt}
The correlation analysis indicates that factors such as electricity market prices, average temperatures across Texas, and ERCOT-wide electricity demand influence the electricity consumption of cryptocurrency mining firms in a complex manner. We observe that these factors can affect each other, necessitating a focus on specific time slots to capture the underlying physics-based relationships. The objective of this section is to perform multivariable linear regression to develop mathematical models describing the electricity consumption of aggregated cryptocurrency mining facilities. We hypothesize the models to be as follows:
\vspace{-10pt}
\begin{align}
E^{M,\text{ns}}_t &= N^{-1}\bigg( \psi^{\text{ns}} T_t \notag \\
&\quad + \mathbb{I}^d(t) \left( \sum_{\forall n \geq 0}\delta^{D,\text{ns}}_n \pi^D_{t-n} + \sum_{\forall n \geq 1} \rho^{D,\text{ns}}_n \pi^R_{t-n} \right) \notag \\
&\quad + \mathbb{I}^p(t) \left( \sum_{\forall n \geq 0} \delta^{P,\text{ns}}_n \pi^D_{t-n} + \sum_{\forall n \geq 1} \rho^{P,\text{ns}}_n \pi^R_{t-n} \right) \notag \\
&\quad \left. + \text{ARMA}^{\text{ns}}(p,d,q)(P,D,Q,[24]) \right)  
\end{align}
\vspace{-10pt}
\begin{align}
E^{M,\text{s}}_t &= N^{-1}\bigg(\psi^{\text{s}} T_t \notag \\
&\quad + \mathbb{I}^d(t) \left( \sum_{\forall n \geq 0} \delta^{D,\text{s}}_n \pi^D_{t-n} + \sum_{\forall n \geq 1} \rho^{D,\text{s}}_n \pi^R_{t-n} \right) \notag \\
&\quad + \mathbb{I}^p(t) \left( \sum_{\forall n \geq 0} \delta^{P,\text{s}}_n \pi^D_{t-n} + \sum_{\forall n \geq 1} \rho^{P,\text{s}}_n \pi^R_{t-n} \right) \notag \\
&\quad + \mathbb{I}^p(t) \sum_{\forall n \geq 1} \gamma_n L_{t-n} \notag \\
&\quad \left. + \text{ARMA}^{\text{s}}(p,d,q)(P,D,Q,[24]) \right) 
\end{align}
\vspace{-2pt}
Here, $E^{M,\text{ns}}_t$ and $E^{M,\text{s}}_t$ are the modeled cryptocurrency mining firms' electricity consumptions during non-summer and summer months. Variable $\psi^{(\cdot)}$ is the regression coefficient for temperature. Parameters $\mathbb{I}^d(t)$, and $\mathbb{I}^p(t)$ are pre-identified binary \sm{indicators}, which equal 1 when the impacts of the associated regressors are active. As discussed earlier, although the market gate closure happens earlier, facilities may still utilize the \sm{day-ahead-market}-cleared data to adjust their bids in the real-time market. The goal here is not to understand how much cryptocurrency miners are bidding into each market but rather to observe how their consumption correlates with historical data, which is why, for day-ahead prices, $n$ can be equal to zero. 

Here, $\delta^{D,\text{s}}_n$ $\rho^{D,\text{s}}_n$ are regression coefficients for price response, and $\delta^{P,\text{s}}_n$ $\rho^{P,\text{s}}_n$ are for peak price response. Cryptocurrency mining facilities might use different predictors, $\gamma_n$, which are combinations of historical ERCOT system-wide electricity demand based on their risk appetite during 4CP hours (4PM - 6PM), also identified through $\mathbb{I}^p(t)$. Finally, the ARMA process models the variance unexplained by the regression model. Here, $N$ is the inverse transformation used to revert the transformed cryptocurrency mining firms' electricity consumption data. Note that all other variables ($T_t$, $\pi^D_{t}$, $\pi^R_{t}$, $L_{t}$) used in this model are also transformed and need to be considered appropriately.

\setlength{\jot}{-2.75pt} 

In this article, contrary to building \sm{the model in a single step}, we perform multiple linear regressions to systematically extract the influence of regressors and perform regression based on the residuals from the previous step. 
To validate the developed regression models, at each step, we divide the data into training and testing samples and compare metrics such as mean squared error (MSE) and root mean squared error (RMSE). Here, we report only statistically significant regressors and their associated p-values for brevity. This section is divided into two \sm{sub}sections dedicated to modeling demand response during non-summer and summer times\sm{, respectively.}

\subsection{Demand response model for the non-summer months}
\vspace{-20pt}

\subsubsection{Temperature effect} 


We initially divided the datasets into training and testing sample days for regression analysis. However, we observed a significant discrepancy in the calculated MSE. We discovered that during certain late nights, the increased temperature led to a decrease in the electricity demand of cryptocurrency mining firms. While we don't know what exactly is responsible for such discrepancy, the removal of days with higher average real-time prices based on their z-scores \sm{addressed} this issue. Using this procedure, even with a 50/50 split of the dataset, the MSE for the training and testing samples remains similar. The calculated correlation coefficient $\psi^{\text{ns}}$ is 0.14 (S.E. = 0.04, p-value = 0.00). We assumed a similar correlation holds during the daytime and removed the associated effect from the dataset to generate the residuals.

\subsubsection{Price effects}


To capture the impact of prices as identified in \eqref{non-summereq}, we regressed the crypto-miners electricity consumption data against electricity price data considering various lag periods. We focused on the hours between 10 AM and 8 PM, setting $\mathbb{I}^d(t) = 1$ during these hours. Our analysis revealed that the strongest p-values occurred when considering day-ahead price data from 2 days prior ($n$=48), real-time prices from the last hour ($n$=1), and day-ahead prices from the previous day ($n$=24). The calculated values are $\delta^{D,\text{ns}}_{48} = -0.08$ (S.E. = 0.03, p-value = 0.01), $\rho^{D,\text{ns}}_{1} = -0.19$ (S.E. = 0.03, p-value = 0.00), and $\rho^{D,\text{ns}}_{24} = -0.11$ (S.E. = 0.03, p-value = 0.00).

Surprisingly, th\sm{ese results} suggest that cryptocurrency mining facilities\sm{,} participating in the day-ahead market\sm{,} may simply observe the most recent publicly available day-ahead prices and \sm{adjust their consumption position accordingly}. One-day-ahead real-time prices are not available at the time of bidding into the day-ahead market. Therefore, despite the higher variability of real-time electricity prices, cryptocurrency miners could be utilizing both one-day-ahead and one-hour-ahead real-time prices to adjust their electricity consumption in the real-time electricity market.

\subsubsection{Peak price effect}

Here, $\mathbb{I}^p(t) = 1$ between 3 PM and 7 PM. Considering a mix of regressors, we found that day-ahead prices from the past hour and real-time prices three hours prior have the strongest correlation. Notably, both sets of data are publicly available for real-time adjustment. The calculated $\delta^{P,\text{ns}}_1$ is -0.16 (S.E. = 0.05, p-value = 0.00), and $\rho^{P,\text{ns}}_3$ is -0.29 (S.E. = 0.05, p-value = 0.00). The negative sign indicates that\sm{,} as cryptocurrency miners observe increasing prices leading to peak hours, they reduce their electricity consumption \sm{monotonically}. Note that this adjustment can only be carried out in the real-time market.



\subsubsection{Autoregressive component}

In the ACF and partial ACF (PACF) factors calculated using the residuals, we observe spikes at lag 1 in the PACF plots without seasonal differencing, implying the strong presence of an AR(1) component. We still observe spikes appearing at around a lag of 24 in both ACF and PACF plots, suggesting the seasonality of the data. Spikes near the lag of 24 in the seasonally differenced PACF data suggest the presence of seasonal autoregressive order. We observe that the ARIMA(1,0,0)(1,1,0)[24] model fits reasonably well based on the Akaike Information Criterion (AIC). The model parameters are given as: $\phi_1$ = 0.83 (S.E. = 0.02, p-value = 0.00), $\Phi_1$ = -0.43 (S.E. = 0.02, p-value = 0.00), $\sigma$ = 0.58 (S.E. = 0.02, p-value = 0.00). The Ljung-Box test shows the lack of autocorrelation in the residuals (p-value = 0.82). The ADF test indicates that the residual is stationary (p-value = 0.82), and the BP test shows that the dataset is weakly heteroskedastic (statistic = 77.5). With a 35/65 split between training and testing samples, we observe an MSE of 1.37, implying that while the ARIMA model captures the variability in the dataset well. Note that these calculations are based on transformed data, and these \sm{figures improve when computed in the original space.}  

\subsubsection{Accuracy of non-summer model}

The empirical equation representing cryptocurrency miners' demand response during the non-summer months is given as:
\vspace{-7pt}
\begin{equation}
\begin{split}
E^{M,\text{ns}}_t &= N^{-1}\left( 0.14 T_t \right. \\
&\quad + \mathbb{I}^d(t)  \left(-0.08 \pi^D_{t-48} -0.19 \pi^R_{t-1} -0.11  \pi^R_{t-24} \right)   \\
&\quad + \mathbb{I}^p(t) \left(-0.16 \pi^D_{t-1} -0.29 \pi^R_{t-3} \right) \\
&\quad \left. + \text{ARMA}^{\text{ns}}(1,0,0)(1,1,0,[24]) \right)
\end{split} \label{non-summereq}
\end{equation}
\vspace{-2pt}
To compute the overall accuracy of the model, we need to compare how much of the variability is explained using correlation analysis alone versus the additional use of an autoregressive model. The mean squared error (MSE) and mean absolute percentage error (MAPE) of the correlation analysis-only model are 25.10 and 3.27\%, respectively. These values change to 32.06 and 3.55\% when using the combined correlation and autoregressive model. However, the true value of the combined model is reflected in the coefficient of determination, which, considering errors only up to the 75\% inter-quantile range, improves from 0.32 to 0.77. An example of a time-series plot comparing true and predicted demand for an arbitrarily selected 7 consecutive days for the non-summer months is provided in Fig. \ref{fig:TimeSeries}(a) (please consider the standard errors described in the paper for the error bound). Our model could not explain a significant amount of variance in the original dataset, which, based on this figure, could be due to the predicted magnitude of peaks.


\subsection{Demand response model for the summer months}
\vspace{-20pt}
\subsubsection{Temperature effect} 

Like non-summer months, we observed similar discrepancies, where during certain late nights, higher temperatures are shown to lead to lower electricity consumption. However, compared to non-summer times, the impact is less prominent here\sm{, which could be due to temperatures remaining high through the summer, thereby masking the relation between temperature and consumption. }
The calculated regression coefficient $\psi^{\text{s}}$ is given as 0.12 (S.E. = 0.04, p-value = 0.01).

\subsubsection{Price effects}

As in the \sm{non-summer} model, we focused on the hours between 10 AM and 8 PM for all four summer months and regressed the electricity consumption data against price data. Here, we observed that the strongest p-values occurred when considering real-time price data from 3 days prior ($n=72$) and the current day-ahead prices ($n=0$). The calculated values are $\delta^{R,\text{s}}_{72} = 0.09$ (S.E. = 0.04, p-value = 0.03) and $\rho^{D,\text{s}}_{0} = -0.40$ (S.E. = 0.04, p-value = 0.00).

This behavior essentially implies that there is a negative correlation between the real-time electricity prices from 3 days prior and the current day-ahead electricity prices. Specifically, if the day-ahead prices are significantly high, the real-time prices will also be higher during the same period, leading cryptocurrency mining firms to \sm{significantly reduce the intensity of} their operations\sm{.}



\subsubsection{Peak price effect}


Prices peak during the summer months, especially in the afternoon hours. These are the same hours when the demands peak as well, and ERCOT calculates 4CP charges based on consumption during these hours. Therefore, it is of interest to isolate how much the cryptocurrency miners are responding because of peak prices from the hedging to avoid 4CP charges. Here, we focus on the former, where we want to investigate, like the non-summer months, how increased prices contribute to cryptocurrency miners' response. Therefore, we focus on July and August datasets, the months with higher price volatility, specifically between 3 PM and 7 PM. We observe that the decision to reduce electricity consumption due to peak electricity prices is \sm{based o}n the recently cleared real-time prices. The coefficient showing the relationship $\rho^{P,\text{s}}_1$ is -0.13 (S.E. = 0.06, p-value = 0.033).


\subsubsection{4CP effect}


Of the 2.9 GW of installed capacity in ERCOT, if some cryptocurrency miners try to avoid the critical peak, the peak demand could shift to later in the same day or even to the next day. To examine these effects, which may occur primarily to avoid 4CP charges, we focus on June and September, specifically between 3 PM and 7 PM. Interestingly, we find that the electricity consumption of cryptocurrency ming firms is a weighted average of their consumption over the past two days during similar hours. The correlation coefficients are given as $\gamma_{24} = -0.89$ (S.E. = 0.11, p-value = 0.00) and $\gamma_{48} = 0.39$ (S.E. = 0.114, p-value = 0.00). This suggests that miners might be basing their behavior on ERCOT's system-wide demand from the previous day. If the demand two days ago was not too high, but the demand yesterday was high, it is likely that today's demand will also be high. This behavior appears to be completely rational.


\subsubsection{Autoregressive component}

Like non-summer months, we observed that the ARMA model (1,0,0)(1,1,1,[24]) could explain a significant part of the variability in the residual dataset. The model parameters are given as: $\phi_1$ = 0.84 (S.E. = 0.01, p-value = 0.00), $\Phi_1$ = -0.09 (S.E. = 0.03, p-value = 0.00),  $\Theta_1$ = -0.93 (S.E. = 0.02, p-value = 0.00) and $\sigma$ = 0.7 (S.E. = 0.01, p-value = 0.00). The Ljung-Box test shows the lack of autocorrelation in the residuals (p-value = 0.88). The ADF test indicates that the residual is stationary (p-value = 0.00), and the BP test shows that the dataset is weakly heteroskedastic (statistic = 94.9). With a 35/65 split between training and testing samples, we observe an MSE of 1.09, implying that while the ARIMA model captures \sm{most of} the variability in the dataset.

\subsubsection{Accuracy of summer model}

The empirical equation representing cryptocurrency miners' demand response during the summer months is given in \sm{\eqref{summereq}}. The residuals suffer from similar issues as was discussed in the earlier model (with RMSE and MAPE of 83.14 and 90.96\% with the correlation-only model to RMSE and MAPE of 60.86 and 64.24\% with the incorporation of autocorrelation); however, the efficacy of the model is further evidenced through the increased coefficient of determination of 0.93 to 0.99, implying that the heuristic-based correlation model itself can explain a significant portion of cryptocurrency miners' behavior, and the model get strengthened with inclusion of ARIMA model.
\vspace{-5pt}
\begin{equation}
\begin{split}
E^{M,\text{s}}_t &= N^{-1}\left( 0.12 T_t \right. \\
&\quad + \mathbb{I}^d(t)  \left(-0.40 \pi^D_{t} +0.09 \pi^R_{t-72} \right)   \\
&\quad + \mathbb{I}^p(t) \left( -0.13 \pi^R_{t-1}\right) \\
&\quad + \mathbb{I}^p(t) \left( -0.89 L_{t-24} + 0.39 L_{t-48}\right) \\
&\quad \left. + \text{ARMA}^{\text{ns}}(1,0,0)(1,1,1,[24]) \right)
\end{split} \label{summereq}
\end{equation}

\begin{figure}
    \centering
      \includegraphics[width=\linewidth]{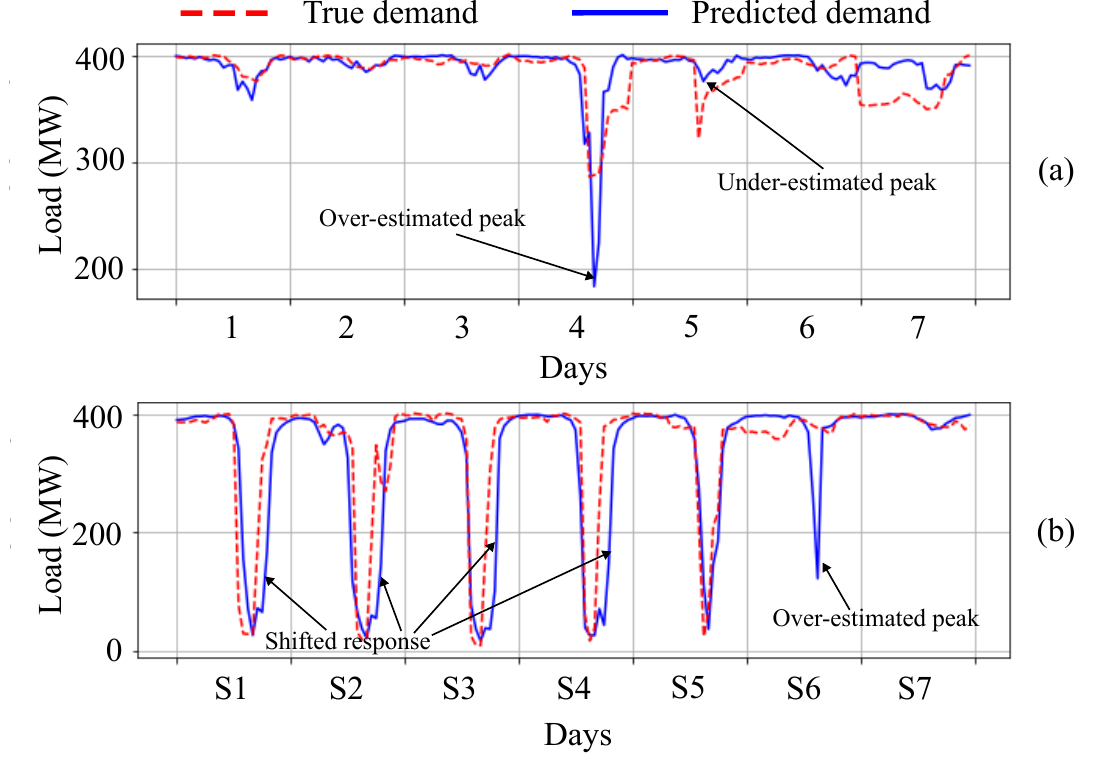}
	\caption{Timeseries of true and predicted electricity demand. (a) for non-summer months, (b) for summer months.}
	\label{fig:TimeSeries}       
\end{figure}

An example of a time-series plot comparing true and predicted demand for an arbitrarily selected 7 consecutive days for the summer months is provided in Fig. \ref{fig:TimeSeries}(b) (please consider the standard errors described in the paper for the error bound).

\vspace{-7pt}
\section{Conclusion}
Through a comprehensive data analysis \sm{we present an} econometric model \sm{that provides} a robust framework for understanding the behavior of large flexible cryptocurrency mining firms in the Texas power grid. By incorporating internal factors through the SARIMA process and external factors via selective \sm{external} correlations, our model achieves reasonable accuracy. The quantile transformation used captures some of the nonlinearities among the variables.  Our analysis reveals that cryptocurrency mining firms' electricity consumption is \sm{mostly} influenced by temperature, electricity prices, and demand response strategies rather than by short-term fluctuations in cryptocurrency prices. This insight highlights the importance of considering multiple factors in predictive modeling.

The practical utility of our model lies in its ability to generate synthetic datasets that can simulate various grid conditions and mining behaviors in Texas. This capability is crucial for power system simulations and for developing strategies to enhance grid reliability and efficiency. Additionally, the developed SARIMA model, with some modifications, could be applied to understand the behavior of mining firms in other regions, particularly where there is a high penetration of cryptocurrency mining and significant interaction with wholesale electricity markets. Furthermore, this study can help power grid operators better anticipate and manage the impact of these emerging technologies on the energy landscape. 

\section{Acknowledgments}
This work is supported in part by Texas A\&M Energy Institute, in part by the U.S. Department of Energy (DOE) project OPEN COG Grid, and in part by the Blockchain and Energy Research Consortium at Texas A\&M University. This work was performed under the auspices of the DOE by Lawrence
Livermore National Laboratory under Contract DE-AC52-07NA27344. LLNL document control number: LLNL-CONF-865441.

\bibliographystyle{apalike}
\bibliography{sample}

\end{document}